\documentclass[[12pt,a4paper,preprint,preprintnumbers,amsmath,amssymb,nofootinbib]{revtex4}
\usepackage{graphicx}
\usepackage{amsmath}
\usepackage{amssymb}
\usepackage{bbold}
\usepackage{type1cm}
\usepackage{rotating}
\usepackage{tabularx}
\usepackage{pdflscape}

\usepackage{dsfont}
\usepackage{overpic}
\usepackage{natbib}
\usepackage{slashed}
\usepackage{subfigure}
\usepackage{bbm}

\allowdisplaybreaks

\begin{document}
\title{A variation on ``compositeness'' (including higher partial waves)}

\author{Peter C.~Bruns}
\affiliation{Nuclear Physics Institute, 25068 \v{R}e\v{z}, Czech Republic }
\date{\today}
\begin{abstract}
The ``spatial interpretation of compositeness'', presented and discussed in \cite{Bruns:2019xgo,Bruns:2022hmb} in the context of non-relativistic potential scattering, is extended to higher partial waves. A particular set of basis states is used to arrive at a slightly different perspective on the derivation and interpretation of ``compositeness'' usually given in the literature.  
 \end{abstract}

\maketitle

\section{Introduction}
\label{sec:Intro}

The concept of compositeness has become widely used in the past two decades to assess the ``nature'' of a given hadronic state, i.~e. to quantify the extent to which it can be considered a ``hadronic molecule'' composed of other hadrons, as opposed to a state that can be directly explained within a quark model (quark-antiquark or three-quark states). This type of analysis can be traced back to Weinberg's study of the case of the deuteron \cite{Weinberg:1965zz}\footnote{See also Sec.~10.7 of \cite{Weinberg:1995mt} for a brief textbook presentation by the same author.}. Basically, the compositeness is given by $1-Z$, where $Z$ is the field renormalization constant for the hadronic state in question, and is sometimes interpreted as ``the probability of finding the bare [``elementary''] state in the bound state'' (see e.g. \cite{Sekihara:2014kya}). A state for which $Z=0$ is then considered as ``fully composite''. Note that we are not talking about compositeness and ``elementariness'' in an absolute sense here: Only particles appearing in asymptotic states coupling to the bound state (or resonance) qualify as possible constituents, whose inner structure is not further resolved. \\
The methods of studying the compositeness of hadrons have meanwhile been generalized in various directions, and have been applied in many interesting case studies \cite{Baru:2003qq,Baru:2010ww,Guo:2017jvc,Matuschek:2020gqe,Hyodo:2008xr,Gamermann:2009uq,Hyodo:2011qc,Aceti:2012dd,Hyodo:2013nka,Nagahiro:2014mba,Guo:2015daa,Sekihara:2016xnq,Oller:2017alp,Li:2021cue,Kinugawa:2024kwb,Kinugawa:2024jwq}.
For a comprehensive presentation of the compositeness concept and its applications, we refer to \cite{Sekihara:2014kya}. See also \cite{vanKolck:2022lqz} for an account from an effective-field-theory perspective, and \cite{Peterken:2023bdv} for a discussion of attempts to define compositeness rigorously in quantum field theories.\footnote{In \cite{Peterken:2023bdv}, doubts are expressed that compositeness can be considered as a rigorous notion in quantum field theory. In parts, the arguments for this conclusion resemble the ones G.~Chew gave for his postulate of ``nuclear democracy'' \cite{Chew:1963zza}, which is based on the constitutional (field-theoretical) right to participate in crossing, and forbids to distinguish elementary from composite strongly-interacting particles.}\\
In \cite{Bruns:2019xgo,Bruns:2022hmb}, we have taken a step back, and limited ourselves to the simplest case in which the above-mentioned concepts could make sense, namely, the non-relativistic scattering of a spinless particle by a spherically symmetric, energy-independent finite-range potential. In close analogy to the relativistic case, we could extract the non-relativistic counterpart of the standardly defined compositeness from the residue of the s-wave bound-state pole of the scattering amplitude, and found that it is proportional to $P(r>d)$, the probability to find the particle at a distance $r$ from the center of the potential greater than the range of the potential $d$. The factor of proportionality is of a universal form, tends to one when $d\rightarrow 0$, and is responsible for the fact that the extracted compositeness can be greater than one. \\
\quad \\
The present contribution has two aims: First, to extend the analysis of \cite{Bruns:2019xgo,Bruns:2022hmb} to bound states of arbitrary orbital angular momentum $\ell$, and second, to mimick the standard quantum-mechanical derivation of the decomposition into compositeness and elementariness (see e.g. \cite{Weinberg:1965zz} and Sec.~2.1. of \cite{Sekihara:2014kya}) with a different set of basis states, derived from a certain limit of an energy-{\em in}dependent potential. The mathematical consequences are essentially the same for each choice of basis states, and thus it should become clear how the relation between the compositeness and a quantity like $P(r>d)$ can be made plausible. \\ 
The set of basis states we use to decompose the bound-state wave functions is explained in Sec.~\ref{sec:HS}, and applied to a bound state in a finite-range potential $V(r)$ in Sec.~\ref{sec:findres}. The chosen basis states are also quite convenient, though not really necessary, for the computation of the residue of the bound-state pole in the scattering amplitude, from which the compositeness (often denoted as $X$) is usually derived in applications (see e.~g. Sec.~3 of \cite{Sekihara:2014kya}).
For $\ell>0$, the relation between the residues of partial-wave amplitudes and the spatial probabilities is, unfortunately, more complicated than in the s-wave case, and is given in Eq.~(\ref{eq:PandCell}) below. Note that this relation is still {\em exact\,} in the present non-relativistic context, and not an approximation for short ranges or particularly small bound-state energies. Such approximations are, however, used to derive relations between the residues and the scattering lengths and effective ranges, as was already done in \cite{Weinberg:1965zz} for the s-wave case. In Sec.~\ref{sec:thrpars}, we give approximate expressions for the spatial probabilities in terms of the threshold parameters, which can be used to get estimates for the range of the potential, and give an impression of the ``size'' of the bound state. We tested those relations for the case of a spherical-well potential, and found the results in reasonable agreement with the derived relations. Our conclusions can be found in Sec.~\ref{sec:conclusions}.

\section{Basis states from a ``hard sphere'' problem}
\label{sec:HS}

Let us divide the universe in an inner and an outer part by a ``delta-shell'' potential \cite{Nussenzveig:1961nc},
\begin{equation}
V_{\mathrm{shell}}(r) = \frac{v}{2R}\delta(r-R)\,.
\end{equation}
We are interested here only in the limit where the strength parameter $v\rightarrow\infty$, so that the sphere of radius $R$ becomes impenetrable. Wave functions for a particle of mass $\mu$ pertaining to this simple case are easily found. In the outer region $r>R$, we have a continuum of states with energies $E>0$, labeled by $k=+\sqrt{2\mu E}$ and angular-momentum quantum numbers $\ell,m$:
\begin{equation}
  \langle\vec{r}|k\ell m\rangle = \Phi_{k\ell m}^{>}(\vec{r}) = \theta(r-R)i^{\ell}\sqrt{\frac{2\mu k}{\pi}}\left(j_{\ell}(kr)+ikf_{\ell}^{\mathrm{h.s.}}(k)h_{\ell}^{+}(kr)\right)\mathcal{Y}_{\ell m}(\theta,\phi)\,,
\end{equation}
where $f_{\ell}^{\mathrm{h.s.}}(k)$ is the partial-wave amplitude for scattering on a hard sphere of radius $R$,
\begin{equation}
f_{\ell}^{\mathrm{h.s.}}(k) = \frac{i}{2k}\left(\frac{h_{\ell}^{-}(kR)}{h_{\ell}^{+}(kR)}+1\right)\,,
\end{equation}
$r:=|\vec{r}|$, $\theta(\cdot)$ is the Heaviside step function, $h^{\pm}_{\ell}(\cdot)$ are the spherical Hankel functions of the first and second kind (see App.~\ref{app:sph_hankel}), $j_{\ell}(z)=\frac{1}{2}\left(h_{\ell}^{+}(z)+h_{\ell}^{-}(z)\right)$ are the spherical Bessel functions of the first kind, and $\mathcal{Y}_{\ell m}(\theta,\phi)$ are the usual spherical harmonics. With the help of the formulae collected in App.~\ref{app:sph_hankel}, one can verify that the $\Phi_{k\ell m}^{>}(\vec{r})$ are orthonormal in the sense
\begin{equation}
\int d^3\vec{r}\,\Phi_{k'\ell' m'}^{>\ast}(\vec{r})\Phi_{k\ell m}^{>}(\vec{r}) = \delta_{\ell'\ell}\delta_{m'm}\frac{\mu}{\sqrt{k'k}}\delta(k'-k) = \delta_{\ell'\ell}\delta_{m'm}\delta(E'-E)\,,
\end{equation}
just like the free wave functions. There is also a discrete set of states, describing particles confined to the inside of the sphere, $r<R$. The corresponding position-space wave functions are 
\begin{equation}\label{eq:nlm_wf}
  \langle\vec{r}|n\ell m\rangle = \Phi_{n\ell m}^{<}(\vec{r}) = \theta(R-r)\mathcal{N}_{n\ell}j_{\ell}\left(\alpha_{n\ell}\frac{r}{R}\right)\mathcal{Y}_{\ell m}(\theta,\phi)\,,\quad \mathcal{N}_{n\ell}=\frac{\sqrt{2/R^3}}{|j_{\ell+1}(\alpha_{n\ell})|}\,.
\end{equation}
Here $\alpha_{n\ell}$ is the $n$th positive zero of the spherical Bessel function $j_{\ell}$. To compute the associated momentum-space wave functions, we employ the expansion of a plane wave into spherical harmonics,
\begin{equation}
\langle\vec{q}|\vec{r}\rangle = \frac{e^{-i\vec{q}\cdot\vec{r}}}{(2\pi)^{3/2}} = \frac{4\pi}{(2\pi)^{3/2}}\sum_{\ell=0}^{\infty}\sum_{m=-\ell}^{\ell}(-i)^{\ell}j_{\ell}(|\vec{q}|r)\mathcal{Y}_{\ell m}(\theta_{q},\phi_{q})\mathcal{Y}_{\ell m}^{\ast}(\theta,\phi)\,, \label{eq:planewavedecomp}
\end{equation}
and the integral of Eq.~(\ref{eq:A10}) to get
\begin{equation}
\langle\vec{q}|n\ell m\rangle = \mathcal{N}_{n\ell}\frac{4\pi(-i)^{\ell}}{(2\pi)^{3/2}}\frac{\alpha_{n\ell}R}{\left(\frac{\alpha_{n\ell}}{R}\right)^2-|\vec{q}|^2}j_{\ell+1}(\alpha_{n\ell})j_{\ell}(|\vec{q}|R)\mathcal{Y}_{\ell m}(\theta_{q},\phi_{q})\,.
\end{equation}
We denote the angles in spherical coordinates pertaining to $\vec{r}$ by $\theta,\phi$, and those pertaining to $\vec{q}$ by $\theta_{q},\phi_{q}$. Note that $\langle\vec{q}|n\ell m\rangle$ is not singular as $|\vec{q}|\rightarrow\alpha_{n\ell}/R$, because $\alpha_{n\ell}$ is a zero of $j_{\ell}\,$. \\
For the continuum states, $\langle\vec{q}|k\ell m\rangle$ could be computed in a similar fashion, employing the integrals in App.~\ref{app:sph_hankel}. As the result is a bit lengthy, and we don't need the explicit result in the following, we shall omit it here. One can use the ``closure relations'' for the spherical Bessel functions and spherical harmonics,
\begin{eqnarray}
  & & \frac{2r'r}{R^3}\sum_{n=1}^{\infty}\frac{j_{\ell}\left(\alpha_{n\ell}\frac{r'}{R}\right)j_{\ell}\left(\alpha_{n\ell}\frac{r}{R}\right)}{|j_{\ell+1}(\alpha_{n\ell})|^2} = \delta(r'-r) \quad \mathrm{for}\,\,0<r,r'<R\,, \label{eq:closure1} \\
  & & \frac{2r'r}{\pi}\int_{0}^{\infty}k^2dk\,j_{\ell}(kr')j_{\ell}(kr) \,\,\,=\, \delta(r'-r) \quad \mathrm{for}\,\,0<r,r'\,, \label{eq:closure2} \\
  & & \sum_{\ell=0}^{\infty}\sum_{m=-\ell}^{\ell}\mathcal{Y}_{\ell m}(\theta',\phi')\mathcal{Y}_{\ell m}^{\ast}(\theta,\phi) = \delta(\cos\theta-\cos\theta')\delta(\phi-\phi')\,, \label{eq:closureY}
\end{eqnarray}
together with the integrals in App.~\ref{app:sph_hankel}, to verify
\begin{eqnarray}
  \sum_{n\ell m}\Phi_{n\ell m}^{<}(\vec{r'})\Phi_{n\ell m}^{<\ast}(\vec{r}) &=& \theta(R-r)\delta^{3}(\vec{r}-\vec{r}\,')\,, \label{eq:comp_nlm}\\
  \int_{0}^{\infty}dE\,\sum_{\ell m}\Phi_{k\ell m}^{>}(\vec{r'})\Phi_{k\ell m}^{>\ast}(\vec{r}) &=& \theta(r-R)\delta^{3}(\vec{r}-\vec{r}\,') \label{eq:comp_klm} \,.
\end{eqnarray}
In this sense, the union of the discrete and the continuous states, $|n\ell m\rangle$ and $|k\ell m\rangle$, forms a complete set of orthonormal states.

\section{Determination of the residues of bound-state poles}
\label{sec:findres}

Let us now forget about the ``hard sphere'', and consider a more general, spherically symmetric finite-range potential $V(r)$. Assume that it has a bound state $|B\rangle$ of energy $E_{B}<0$, and angular-momentum quantum numbers $\ell,m$, $|B\rangle \equiv |E_{B}\ell m\rangle$. To this bound state corresponds a pole term of the form
\begin{equation}
\mathcal{T}(\vec{q'},\vec{q};E) = \frac{\langle\vec{q'}|\hat{V}|B\rangle\langle B|\hat{V}|\vec{q}\rangle}{E-E_{B}} + \ldots
\end{equation}
in the scattering amplitude $\mathcal{T}$ for the considered potential, with incoming and outgoing momenta $\vec{q}$ and $\vec{q'}$, respectively, and energy $E=k^2/(2\mu)$ (see e.g. Sec.~II of \cite{Bruns:2019xgo}). Here $\hat{V}$ is the operator associated with $V(r)$. On-shell partial-wave scattering amplitudes $f_{\ell}(E)$ are derived from $\mathcal{T}(\vec{q'},\vec{q};E)$ via
\begin{equation}\label{def:fell}
\mathcal{T}_{\ell}(E):=\frac{1}{2}\int_{-1}^{1}dz\,\mathcal{P}_{\ell}(z)\,\mathrm{lim}_{|\vec{q}\,'|,|\vec{q}|\rightarrow k}\mathcal{T}(\vec{q}\,',\vec{q};E)\,,\qquad f_{\ell}(E):=-(2\pi)^2\mu\mathcal{T}_{\ell}(E)\,.
\end{equation}
Here $z$ is the cosine of the scattering angle (the angle between $\vec{q}\,'$ and $\vec{q}$), and $\mathcal{P}_{\ell}(z)$ denote the usual Legendre polynomials. The conventionally normalized partial-wave amplitudes $f_{\ell}(E)$ obey the unitarity requirement $\mathrm{Im}\,(f_{\ell}(E))^{-1}=-k\,$ for real $E>0$, and can conveniently be written as
\begin{equation}\label{eq:fK}
f_{\ell}(E) = \lbrack K_{\ell}^{-1}(E)-ik\rbrack^{-1}\,,
\end{equation}
where $K_{\ell}(E)$ is real for real energies $E$. We see that, to find the residue of the bound-state pole in $f_{\ell}$, we have to consider the matrix elements
\begin{equation}
\langle\vec{q}|\hat{V}|B\rangle = \left(E_{B}-\frac{q^2}{2\mu}\right)\langle\vec{q}|B\rangle \,, \qquad q := \sqrt{(\vec{q}\cdot\vec{q})}\,,
\end{equation}
in the limit where $q\rightarrow k\rightarrow i\kappa_{B} \equiv i\sqrt{-2\mu E_{B}}\,$, $\kappa_{B}>0\,$. This means we have to extract the residue of the pole of $\langle\vec{q}|B\rangle$ at $q^2=2\mu E_{B}$ (we will soon see that $\langle\vec{q}|B\rangle$ indeed has such a pole). To do this, we expand $|B\rangle$ in the set of states considered in Sec.~\ref{sec:HS}:
\begin{equation}
|B\rangle = \sum_{n=1}^{\infty}\langle n\ell m|B\rangle |n\ell m\rangle + \int_{0}^{\infty}dE\,\langle k\ell m|B\rangle |k\ell m\rangle\,.
\end{equation}
In the usual manner of speaking, one could call $\sum_{n}|\langle n\ell m|B\rangle|^2$ ``the probability to find a state confined to a sphere of radius $R$ in the bound state'', etc. However, it is clear from the construction in Sec.~\ref{sec:HS} that this number is just the probability to find the bound particle at a distance $r\leq R$ from the center of the potential. Clearly,
\begin{equation}\label{eq:decomp1}
1=\langle B|B\rangle = \sum_{n}|\langle n\ell m|B\rangle|^2 + \int_{0}^{\infty}dE\,|\langle k\ell m|B\rangle|^2\,.
\end{equation}

Let the finite range of the potential $V(r)$ be $d$. In the following, we will always take the number $R$ associated with the ``hard sphere'' states to satisfy $R>d$. \\

Using again Eqs.~(\ref{eq:planewavedecomp}), (\ref{eq:closure1}) and (\ref{eq:A10}), we can compute the sum (for fixed $\ell,m$, $|m|\leq\ell$)
\begin{eqnarray}
 & & \,\sum_{n=1}^{\infty}\langle\vec{q}|n\ell m\rangle \langle n\ell m|\vec{q}\rangle \nonumber \\ &=& \,\frac{(4\pi)^2}{(2\pi)^3}\frac{R^3}{2}\left((j_{\ell}(qR))^2 + (j_{\ell+1}(qR))^2 - \frac{2\ell+1}{qR}j_{\ell}(qR)j_{\ell+1}(qR)\right)\left|\mathcal{Y}_{\ell m}(\theta_{q},\phi_{q})\right|^2.
\end{eqnarray}
Since also $\sum_{n=1}^{\infty}\langle B|n\ell m\rangle \langle n\ell m|B\rangle \leq 1$, we have (by the Cauchy-Schwarz inequality for series)
\begin{eqnarray*}
 & & \,\left|\sum_{n=1}^{\infty}\langle\vec{q}|n\ell m\rangle \langle n\ell m|B\rangle\right|  \\ &\leq& \,\frac{4\pi}{(2\pi)^{3/2}}\sqrt{\frac{R^3}{2}}\left((j_{\ell}(qR))^2 + (j_{\ell+1}(qR))^2 - \frac{2\ell+1}{qR}j_{\ell}(qR)j_{\ell+1}(qR)\right)^{\frac{1}{2}}\left|\mathcal{Y}_{\ell m}(\theta_{q},\phi_{q})\right|\,,
\end{eqnarray*}
which is clearly bounded for any fixed $R>0$ when $q^2\rightarrow 2\mu E_{B}$, and therefore the pole in $\langle\vec{q}|B\rangle$ can only be produced by the long-range part, $\int_{0}^{\infty}dE\,\langle\vec{q}|k\ell m\rangle \langle k\ell m|B\rangle$\,.
This can be computed explicitly, because we chose $R>d$, and the bound-state wave functions for $r>d$ are of the universal form (with some choice of phase)
\begin{equation}\label{eq:Bslongdist}
\langle\vec{r}|B\rangle\bigl|_{r>d} = i^{\ell+2}\kappa_{B}\mathcal{N}_{B}h_{\ell}^{+}(i\kappa_{B}r)\mathcal{Y}_{\ell m}(\theta,\phi)\quad\overset{r\rightarrow\infty}{\longrightarrow}\quad \mathcal{N}_{B}\frac{e^{-\kappa_{B}r}}{r}\mathcal{Y}_{\ell m}(\theta,\phi)\,,
\end{equation}
where the normalization constant $\mathcal{N}_{B}$ depends on the potential $V$ (as does $\kappa_{B}$, of course)\footnote{We do not add a label $\ell m$ to quantities like $\kappa_{B}$, $\mathcal{N}_{B}$ or $P(r>R)$ for better readability.}. \\
We can use the completeness of the $\Phi_{k\ell m}^{>}$ in the ``outer region'' $r>R$, Eq.~(\ref{eq:comp_klm}), in the form
\begin{equation*}
\int_{0}^{\infty}dE\,\langle\vec{r}\,'|k\ell m\rangle \langle k\ell m|\vec{r}\rangle = \theta(r'-R)\theta(r-R)\frac{1}{r'r}\delta(r'-r)\mathcal{Y}_{\ell m}(\theta',\phi')\mathcal{Y}_{\ell m}^{\ast}(\theta,\phi)\,,
\end{equation*}
to obtain, via Eq.~(\ref{eq:int_hh}),
\begin{eqnarray}
  & & \int_{0}^{\infty}dE\,\langle\vec{q}|k\ell m\rangle \langle k\ell m|B\rangle = -\frac{4\pi}{(2\pi)^{3/2}}\mathcal{Y}_{\ell m}(\theta_{q},\phi_{q})\kappa_{B}\mathcal{N}_{B}\int_{R}^{\infty}r^2dr\,j_{\ell}(qr)h_{\ell}^{+}(i\kappa_{B}r) \nonumber \\
  &=& \frac{4\pi}{(2\pi)^{3/2}}\mathcal{Y}_{\ell m}(\theta_{q},\phi_{q})\left(\frac{\kappa_{B}\mathcal{N}_{B}R^2}{q^2+\kappa_{B}^2}\right)\left(qh_{\ell}^{+}(i\kappa_{B}R)j_{\ell+1}(qR) - i\kappa_{B}j_{\ell}(qR)h_{\ell+1}^{+}(i\kappa_{B}R)\right)\,.
\end{eqnarray}
The complex conjugate of the last expression, $\int_{0}^{\infty}dE\,\langle B|k\ell m\rangle \langle k\ell m|\vec{q}\rangle$, picks up an additional factor $(-1)^{\ell}$ because $(h_{\ell}^{+}(i\kappa_{B}R))^{\ast} = (-1)^{\ell}h_{\ell}^{+}(i\kappa_{B}R)$\,.  
Using the identity (\ref{eq:hankel_id}), this yields 
\begin{equation}
\int_{0}^{\infty}dE\,\langle\vec{q}|k\ell m\rangle \langle k\ell m|B\rangle = \mathcal{N}_{B}\frac{4\pi}{(2\pi)^{3/2}}\mathcal{Y}_{\ell m}(\theta_{q},\phi_{q})\left(\frac{1}{q^2+\kappa_{B}^2} + \dots\right)\,,
\end{equation}
where the dots indicate terms which are regular as $q\rightarrow i\kappa_{B}$. All in all we find
\begin{equation}
 \mathrm{lim}_{q\rightarrow i\kappa_{B}}\left(E_{B}-\frac{q^2}{2\mu}\right)\langle\vec{q}|B\rangle   =  -\frac{\mathcal{N}_{B}}{2\mu}\frac{4\pi}{(2\pi)^{3/2}}\mathcal{Y}_{\ell m}(\theta_{q},\phi_{q})\,.
\end{equation}
For the on-shell partial-wave amplitudes (see Eq.~(\ref{def:fell})), seen as functions of the energy $E$, this means that the residue at the bound-state pole on the negative energy axis is given by
\begin{eqnarray}
  \mathrm{Res}_{E_{B}}f_{\ell} &=& - (-1)^{\ell}\frac{\mathcal{N}_{B}^2}{2\mu}\,,\quad \mathcal{C}_{B}^{\ell} := -\frac{\mu}{\kappa_{B}}(-1)^{\ell}\mathrm{Res}_{E_{B}}f_{\ell}\,,\label{eq:PandCell}\\
  P(r>R) &=&(\kappa_{B}R)^3\biggl(|h_{\ell+1}^{+}(i\kappa_{B} R)|^2-|h_{\ell}^{+}(i\kappa_{B} R)|^2 - \frac{2\ell+1}{\kappa_{B} R}|h_{\ell}^{+}(i\kappa_{B} R)h_{\ell+1}^{+}(i\kappa_{B} R)|\biggr)\mathcal{C}_{B}^{\ell}\nonumber\,.
\end{eqnarray}
For $\ell=0$, this yields
\begin{equation}\label{eq:PandC0}
  P(r>R) = e^{-2\kappa_{B}R}\mathcal{C}_{B}^{0}=:p_{0}(\kappa_{B}R)\mathcal{C}_{B}^{0}\,,
\end{equation}
in agreement with \cite{Bruns:2019xgo}, while for $\ell>0$, we write (\ref{eq:PandCell}) as
\begin{equation}\label{eq:PandCell2}
\mathcal{C}_{B}^{\ell} = (\kappa_{B}R)^{2\ell-1}P(r>R)\,/\,p_{\ell}(\kappa_{B}R)\,.
\end{equation}
For example, for the cases $\ell=1$ to $\ell=3$,
\begin{eqnarray*}
  p_{1}(\kappa R) &=& e^{-2\kappa R}(2+\kappa R)\,,\\
  p_{2}(\kappa R) &=& e^{-2\kappa R}(6+\kappa R(12+\kappa R(6+\kappa R)))\,,\\
  p_{3}(\kappa R) &=& e^{-2\kappa R}(90 + 180\kappa R + 150 (\kappa R)^2 + 60 (\kappa R)^3 + 12 (\kappa R)^4 + (\kappa R)^5)\,.
\end{eqnarray*}

\section{Threshold parameters}
\label{sec:thrpars}

Considering the function $k^{2\ell}(K_{\ell}(E))^{-1}$, and comparing the expansions around $E=0$ and a small $E_{B}<0$, respectively,
\begin{eqnarray}
  k^{2\ell}(K_{\ell}(E))^{-1} &=& -\frac{1}{a_{\ell}} + \mu E r_{\ell} + \ldots\,,\label{eq:ERE} \\
  k^{2\ell}(K_{\ell}(E))^{-1} &=& (i\kappa_{B})^{2\ell}(K_{\ell}(E_{B}))^{-1} + (E-E_{B})\frac{dk^{2\ell}K_{\ell}^{-1}}{dE}\biggr|_{E_{B}} + \ldots \,,
\end{eqnarray}
and using that $(K_{\ell}(E_{B}))^{-1}+\kappa_{B}=0$ and (from the computation of the residue of $f_{\ell}(E)$)
\begin{equation}
\frac{dK_{\ell}^{-1}}{dE}\biggr|_{E_{B}} = \frac{\mu}{\kappa_{B}}\left(1-\frac{(-1)^{\ell}}{\mathcal{C}_{B}^{\ell}}\right)\,,
\end{equation}
we expect, when $\kappa_{B}$ is sufficiently small, 
\begin{equation*}
\frac{1}{a_{\ell}} \approx (-1)^{\ell}\kappa_{B}^{2\ell+1} + E_{B} \frac{dk^{2\ell}K_{\ell}^{-1}}{dE}\biggr|_{E_{B}}\,,\qquad \mu r_{\ell} \approx \frac{dk^{2\ell}K_{\ell}^{-1}}{dE}\biggr|_{E_{B}}\,,
\end{equation*}
as long as the higher terms in the effective-range expansion (\ref{eq:ERE}) do not blow up as $\kappa_{B}\rightarrow 0$ compared to the terms involving $a_{\ell}$ and $r_{\ell}$, so that the first two terms dominate this expansion. This assumption leads us to the approximations
\begin{equation}
\kappa_{B}a_{\ell} \approx \frac{2\mathcal{C}_{B}^{\ell}}{\kappa_{B}^{2\ell}(1+(-1)^{\ell}\mathcal{C}_{B}^{\ell}(1-2\ell))}\,,\qquad  \kappa_{B}r_{\ell} \approx -\kappa_{B}^{2\ell}\left(\frac{1}{\mathcal{C}_{B}^{\ell}} -(-1)^{\ell}(2\ell+1)\right)\,.
\end{equation}
For $\ell=0$, we get Weinberg's relations \cite{Weinberg:1965zz}, identifying $\mathcal{C}_{B}^{0}=1-Z=:X$, while for $\ell>0$, we use Eq.~(\ref{eq:PandCell2}) to write 
\begin{eqnarray}
  \kappa_{B}^2a_{\ell} &\approx& \frac{2R^{2\ell-1}P(r>R)}{p_{\ell}(\kappa_{B}R) - (2\ell-1)(-1)^{\ell}(\kappa_{B}R)^{2\ell-1}P(r>R)}\,,\nonumber \\
  r_{\ell} &\approx& -\frac{1}{R^{2\ell-1}}\left(\frac{p_{\ell}(\kappa_{B}R)}{P(r>R)}-(-1)^{\ell}(2\ell+1)(\kappa_{B}R)^{2\ell-1}\right)\,, \quad\mathrm{or}\quad\nonumber \\
  P(r>R) &\approx& \frac{\kappa_{B}^2 a_{\ell}p_{\ell}(\kappa_{B} R)}{R^{2\ell-1}\left(2+(2\ell-1)(-1)^{\ell}\kappa_{B}^{2\ell+1}a_{\ell}\right)} =: P_{a}(r>R)\,, \label{eq:Pa_approx} \\
  P(r>R) &\approx& \frac{p_{\ell}(\kappa_{B}R)}{R^{2\ell-1}\left((2\ell+1)(-1)^{\ell}\kappa_{B}^{2\ell-1} - r_{\ell}\right)}  \,\,\,=: P_{r}(r>R)\,. \label{eq:Pr_approx}
\end{eqnarray}
\quad \\
\underline{{\bf Example: Spherical well}}

As an example, we consider bound states in a spherical-well potential (see App.~\ref{app:sph_well}) of range $d=5\,\mu^{-1}$, $V(r)=V_{0}\theta(d-r)$. For each $\ell$, we adjust $V_{0}$ so that a bound state at $\kappa_{B}=0.1\,\mu$ emerges. In the following, all numbers are given in the appropriate units of $\mu=1$ (and using particle physics units, $\hbar=c=1$). The table below displays the according numbers for $\ell=0,1,2,3\,$. We have checked that the relations in Eqs.~(\ref{eq:PandCell})-(\ref{eq:PandCell2}) are exactly fulfilled. One sees that the approximations $P_{a,r}$ for $P(r>d)$ work very well for $\ell=0$, and are a few percent off for $\ell>0$ (the $\ell=0$ case is numerically similar to the case of the deuteron, see also App.~B of \cite{Bruns:2022hmb}). We refer to \cite{Kinugawa:2021ykv,Song:2022yvz,Albaladejo:2022sux,Kinugawa:2022fzn,Yin:2023wls} for more in-depth discussions of the role of the interaction range, and the case of the deuteron in particular.  \\

\begin{figure}[h]
\centering
\subfigure[\,$\ell=0$]{\includegraphics[width=0.45\textwidth]{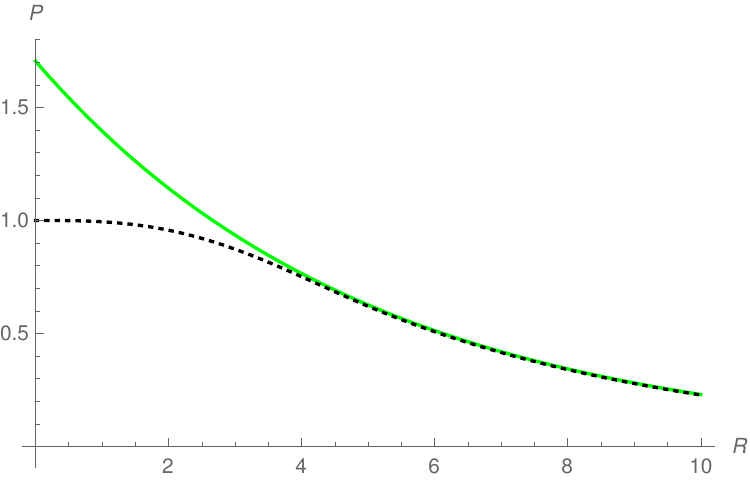}}
\subfigure[\,$\ell=1$]{\includegraphics[width=0.45\textwidth]{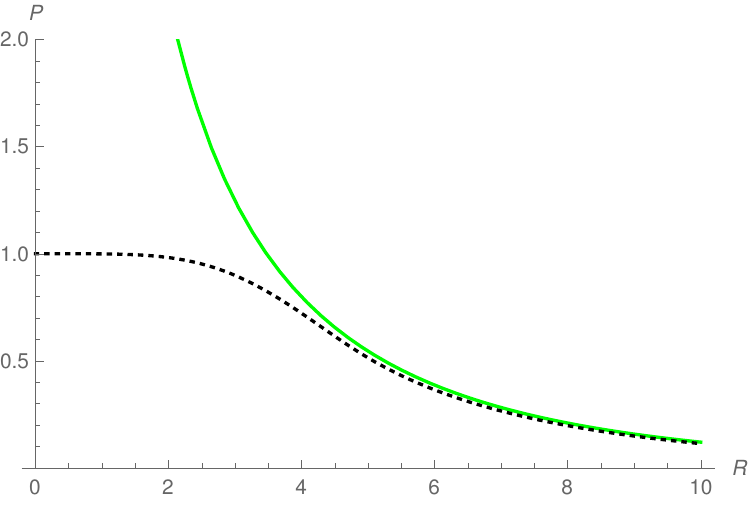}}
\subfigure[\,$\ell=2$]{\includegraphics[width=0.45\textwidth]{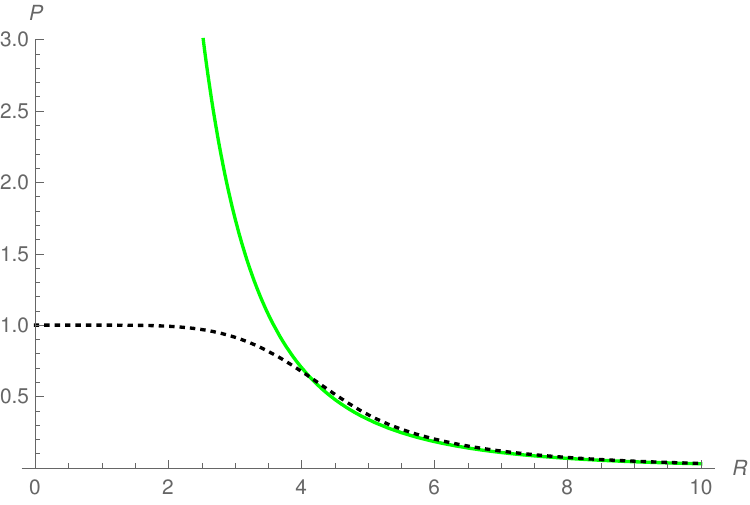}}
\subfigure[\,$\ell=3$]{\includegraphics[width=0.45\textwidth]{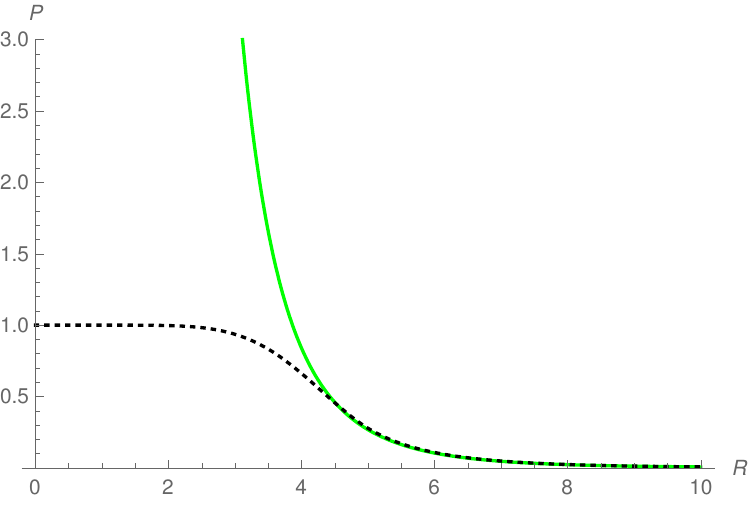}}
\caption{The estimate $P_{a}(r>R)$ (green lines) for the true $P(r>R)$ (black dashed lines) for $d=5$, $\mu=1$, $\kappa_{B}=0.1\,$, $\ell=0,1,2,3\,$. The estimate $P_{r}$ looks very similar to $P_{a}$, so it is not shown here.}
\label{fig:PR}
\end{figure}

\begin{center}
  \begin{tabular}{|c|c|c|c|c|c|c|c|}
  \hline
  $\ell$\quad & $V_{0}$ & $\mathcal{C}^{\ell}_{B}$ & $a_{\ell}$ & $r_{\ell}$ & $P(r>d)$ & $P_{a}(r>d)$ & $P_{r}(r>d)$ \\
  \hline
  \,0\, & \, -0.0725 \qquad &  \, 1.6885 \quad & \, 12.589  \quad & \, 4.1893  \quad  & \, 0.6203 \quad & \, 0.6265\quad & \, 0.6328 \quad \\
  \,1\, & \, -0.2089 \qquad &  \, 0.2794 \quad & \, 456.950 \quad & \, -0.6167 \quad  & \, 0.5137 \quad & \, 0.5451\quad & \, 0.5808 \quad \\
  \,2\, & \, -0.4119 \qquad &  \, 0.0092 \quad & \, 1743.991\quad & \, -0.1236 \quad  & \, 0.3721 \quad & \, 0.3393\quad & \, 0.3123 \quad \\
  \,3\, & \, -0.6713 \qquad &  \, 0.0001 \quad & \, 2011.072\quad & \, -0.1034 \quad  & \, 0.2774 \quad & \, 0.2671\quad & \, 0.2572 \quad \\
  \hline
\end{tabular}
\end{center}
\quad \\
In practice, one might not know the form of the potential, but only the numbers $a_{\ell},\,r_{\ell}$ and $\kappa_{B}$. Still, we can gain some insight into the spatial structure of the bound state and the potential from $P_{a,r}(r>R)$ of Eqs.~(\ref{eq:Pa_approx}), (\ref{eq:Pr_approx}). To demonstrate this, we plot in Fig.~\ref{fig:PR}  $P_{a}(r>R)$ over $R$ (green lines) for $\ell=0,1,2,3\,$. It is clear that this approximation to the true $P(r>R)$ (black dashed lines) cannot make sense when $P_{a}(r>R)>1$. This happens when $R$ is markedly smaller than the interaction range $d$, contrary to the assumption $R>d$ we made in our derivations. So, from the green lines alone, we can already estimate the range of the potential to be $d\gtrsim 4\,\mu^{-1}$. In the region $R>d$, where our derivations are valid, $P_{a,r}(r>R)$ both yield a decent estimate for the true $P(r>R)$ as computed from the exact wave functions. \\

\section{Conclusions}
\label{sec:conclusions}

In Eq.~(\ref{eq:decomp1}), the states $|n\ell m\rangle$ play the role of the ``elementary'' states in the usual derivations of compositeness, while the role of the ``free two-particle states'' is played by the continuum of hard-sphere scattering states $|k\ell m\rangle$. It is clear from (\ref{eq:nlm_wf}), (\ref{eq:PandCell}) and the computation of the residues in Sec.~\ref{sec:findres} that $\mathrm{Res}_{E_{B}}f_{\ell}$ and $\mathcal{C}^{\ell}_{B}$ tend to zero when the bound state becomes confined to a small region $r\leq d$ (compare also the numerical examples in the tables in \cite{Bruns:2022hmb}). \\
Under the assumption that the first two terms in the effective-range expansion still dominate when $\kappa_{B}\rightarrow 0$, we can use the {\em exact\,} relations in Eq.~(\ref{eq:PandCell}) to get approximate relations, valid for sufficiently small $\kappa_{B}$, that connect the threshold parameters with the spatial probability distribution of the bound state. We found that those relations are reasonably well satisfied for a spherical-well potential when $\kappa_{B}=0.1\,\mu$ and $R>d=5\,\mu^{-1}$. Moreover, in the case $\ell=0$, they reduce to the relations already given in \cite{Weinberg:1965zz} if one identifies our $\mathcal{C}_{B}^{0}$, as derived from the bound-state residue of the s-wave scattering amplitude, with $X:=1-Z$. \\
Obviously, the present contribution is limited to the simple case of nonrelativistic potential scattering, and relativistic dynamics, energy-dependent potentials, coupled channels and resonances must be studied with more elaborate methods.

\clearpage

\begin{appendix}

\section{Spherical Hankel functions}
\label{app:sph_hankel}
\def\theequation{\Alph{section}.\arabic{equation}}
\setcounter{equation}{0}

The spherical Hankel functions $h_{\ell}^{\pm}(z)$ can be represented by the series
\begin{eqnarray}
  h_{\ell}^{+}(z) &=& (-i)\frac{e^{i(z-\ell\frac{\pi}{2})}}{z}\sum_{k=0}^{\ell}\frac{(+i)^{k}a_{k}(\ell)}{z^{k}}\,,\quad 
  h_{\ell}^{-}(z) = (+i)\frac{e^{-i(z-\ell\frac{\pi}{2})}}{z}\sum_{k=0}^{\ell}\frac{(-i)^{k}a_{k}(\ell)}{z^{k}}\,,\\
  a_{0\leq k\leq\ell}(\ell) &=& \frac{1}{2^{k}k!}\frac{\Gamma(\ell+1+k)}{\Gamma(\ell+1-k)}\,,\quad a_{k>\ell}(\ell)=0\,.
\end{eqnarray}
They are related to the spherical Bessel functions of the first and second kind, $j_{\ell}(z)$ and $n_{\ell}(z)$, by $h_{\ell}^{\pm}=j_{\ell}\pm in_{\ell}\,$. We have the recursion relation
\begin{equation}\label{eq:hankel_red}
h^{\pm}_{\ell+1}(z) = \frac{2\ell+1}{z}h^{\pm}_{\ell}(z)-h^{\pm}_{\ell-1}(z)\,,
\end{equation}
the expression for the derivative,
\begin{equation}\label{eq:hankel_der}
(h_{\ell}^{\pm})'(z)=(\ell/z)h^{\pm}_{\ell}(z)-h^{\pm}_{\ell+1}(z)\,,
\end{equation}
the asymptotic behavior
\begin{equation}\label{eq:hankel_asymp}
h_{\ell}^{\pm}(z)\,\overset{z\rightarrow\infty}{\longrightarrow}\,(\mp i)^{\ell+1}\frac{e^{\pm iz}}{z}\,,
\end{equation}
and the identity
\begin{equation}\label{eq:hankel_id}
h_{\ell}^{+}(z)h_{\ell+1}^{-}(z) - h_{\ell}^{-}(z)h_{\ell+1}^{+}(z) = \frac{2i}{z^2}\,.
\end{equation}
\quad \\
The following integrals are very useful for the investigation in this work:
\begin{eqnarray}
  \int_{R}^{\infty}dr\,r^2h_{\ell}^{+}(i\kappa'r)h_{\ell}^{+}(i\kappa r) &=&  \frac{\kappa\kappa'R^3}{\kappa^2-\kappa'^2}\left(\frac{1}{i\kappa R}h_{\ell}^{+}(i\kappa R)h_{\ell+1}^{+}(i\kappa' R) - \frac{1}{i\kappa' R}h_{\ell}^{+}(i\kappa' R)h_{\ell+1}^{+}(i\kappa R)\right)\,,\nonumber \\
  \int_{R}^{\infty}dr\,r^2h_{\ell}^{\pm}(\alpha r)h_{\ell}^{+}(i\kappa r) &=&  \frac{i\alpha\kappa R^3}{\alpha^2+\kappa^2}\left(\frac{1}{\alpha R}h_{\ell}^{\pm}(\alpha R)h_{\ell+1}^{+}(i\kappa R) - \frac{1}{i\kappa R}h_{\ell}^{+}(i\kappa R)h_{\ell+1}^{\pm}(\alpha R)\right)\,,\label{eq:int_hh}
\end{eqnarray}
where $\kappa',\,\kappa>0$, $R>0,\,\alpha>0$. In the limit $\kappa'\rightarrow\kappa$, the first integral yields
\begin{equation}\label{eq:hankelnormint}
 \int_{R}^{\infty}dr\,r^2|h_{\ell}^{+}(i\kappa r)|^2 = \frac{R^3}{2}\left(|h_{\ell+1}^{+}(i\kappa R)|^2-|h_{\ell}^{+}(i\kappa R)|^2 - \frac{2\ell+1}{\kappa R}|h_{\ell}^{+}(i\kappa R)h_{\ell+1}^{+}(i\kappa R)|\right)\,,
\end{equation}
using Eq.~(\ref{eq:hankel_der})\,. Note that $h_{\ell}^{+}(i\kappa R) = -(-i)^{\ell}|h_{\ell}^{+}(i\kappa R)|\,$. For small $\kappa R$, this behaves as
\begin{eqnarray*}
  \int_{R}^{\infty}dr\,r^2|h_{\ell}^{+}(i\kappa r)|^2 &=& \frac{R^3}{2}\frac{1}{(\kappa R)^3}\left(1+\mathcal{O}(\kappa R)\right)\quad\mathrm{for}\quad \ell=0\,,\\
  \int_{R}^{\infty}dr\,r^2|h_{\ell}^{+}(i\kappa r)|^2 &=& \frac{R^3}{(\kappa R)^{2\ell+2}}\biggl(\frac{\Gamma(2\ell-1)\Gamma(2\ell)}{4^{\ell-1}(\Gamma(\ell))^2} +\mathcal{O}(\kappa R)\biggr)\quad\mathrm{for}\quad \ell>0\,.
\end{eqnarray*}
\newpage
We also need some integrals involving spherical Bessel functions of the first kind, e.g.
\begin{eqnarray}
  \int_{R}^{\infty}dr\,r^2j_{\ell}(k'r)j_{\ell}(kr) &=& \frac{\pi}{2kk'}\delta(k'-k) - \frac{R^2}{k^2-k'^2}\left(k j_{\ell+1}(kR)j_{\ell}(k'R) - k' j_{\ell+1}(k'R)j_{\ell}(kR)\right)  \nonumber \\
  &\rightarrow& \frac{\pi}{2kk'}\delta(k'-k) \quad\mathrm{for}\,R\rightarrow 0\,, \label{eq:A9}\\
  \int_{0}^{R}dr\,r^2j_{\ell}(k'r)j_{\ell}(kr) &=& \frac{R^2}{k^2-k'^2}\left(k j_{\ell+1}(kR)j_{\ell}(k'R) - k' j_{\ell+1}(k'R)j_{\ell}(kR)\right) \label{eq:A10}  \\
  &\rightarrow& \frac{R^3}{2}\left((j_{\ell}(kR))^2 + (j_{\ell+1}(kR))^2 - \frac{2\ell+1}{kR}j_{\ell}(kR)j_{\ell+1}(kR)\right) \quad\mathrm{for}\,\,k'\rightarrow k\,,\nonumber 
\end{eqnarray}
for $k',\,k >0$ and $R>0\,$. Similarly,
\begin{eqnarray}
  \int_{R}^{\infty}dr\,r^2n_{\ell}(k'r)n_{\ell}(kr) &=& \frac{\pi}{2kk'}\delta(k'-k) - \frac{R^2}{k^2-k'^2}\left(k n_{\ell+1}(kR)n_{\ell}(k'R) - k' n_{\ell+1}(k'R)n_{\ell}(kR)\right) \,, \nonumber \\
  \int_{0}^{R}dr\,r^2n_{\ell}(k'r)j_{\ell}(kr) &=&  \frac{R^2}{k^2-k'^2}\left(k j_{\ell+1}(kR)n_{\ell}(k'R) - k' n_{\ell+1}(k'R)j_{\ell}(kR)\right) - \frac{\frac{k^{\ell}}{k'^{\ell+1}}}{k^2-k'^2}\,,\nonumber \\
  \int_{R}^{\infty}dr\,r^2n_{\ell}(k'r)j_{\ell}(kr) &=& \frac{1}{2kk'}\frac{1}{k-k'} - \frac{R^2}{k^2-k'^2}\left(k j_{\ell+1}(kR)n_{\ell}(k'R) - k' n_{\ell+1}(k'R)j_{\ell}(kR)\right) \nonumber \\ &-& \frac{1}{2kk'}\mathcal{P}\left(\frac{1}{k-k'}\right)\,,\label{eq:besselnjintegral}
\end{eqnarray}
where $\mathcal{P}(\cdot)$ denotes the ``principal value'' distribution,
\begin{equation*}
\int_{-b}^{c}dx\,\mathcal{P}\left(\frac{1}{x}\right)f(x) = \mathrm{lim}_{\varepsilon\rightarrow 0}\left(\int_{-b}^{-\varepsilon}dx\,\frac{f(x)}{x} + \int_{\varepsilon}^{c}dx\,\frac{f(x)}{x}\right)\,.
\end{equation*}
Note that the first line in the last Eq.~(\ref{eq:besselnjintegral}) is non-singular for $k'\rightarrow k$, as can be seen with the help of (\ref{eq:hankel_id}). We also dropped terms $\sim\delta(k'+k)$, since we always assume $k',k>0$ here. 

\newpage

\section{Spherical well}
\label{app:sph_well}
\def\theequation{\Alph{section}.\arabic{equation}}
\setcounter{equation}{0}

Here the potential is taken as $V(r)=V_{0}\theta(d-r)$\,, so that the expansion of the scattering amplitude in the coupling strength $V_{0}<0$ starts as
\begin{eqnarray}
  \mathcal{T}(\vec{q}\,',\vec{q};E) &=& V(\vec{q}\,',\vec{q}) + \ldots\,,\quad\mathrm{where} \\
  V(\vec{q}\,',\vec{q}) &=& \int\frac{d^3r}{(2\pi)^3}e^{-i(\vec{q}\,'-\vec{q})\cdot\vec{r}}V_{0}\theta(d-r) \nonumber \\
  &=& \left(\frac{V_{0}}{2\pi^2}\right)\frac{\sin(|\vec{q}\,'-\vec{q}|d)-|\vec{q}\,'-\vec{q}|d\cos(|\vec{q}\,'-\vec{q}|d)}{|\vec{q}\,'-\vec{q}|^3}\,.
\end{eqnarray}
The partial-wave amplitudes for scattering on this potential are given by Eq.~(\ref{eq:fK}), with
\begin{equation}
K_{\ell}(E) = \frac{1}{k}\,\frac{k j_{\ell}(\xi d)j'_{\ell}(k d) - \xi j'_{\ell}(\xi d)j_{\ell}(k d)}{k j_{\ell}(\xi d)n'_{\ell}(k d) - \xi j'_{\ell}(\xi d)n_{\ell}(k d)}\,
\end{equation}
(see App.~\ref{app:sph_hankel} for the spherical Bessel functions), where $k=+\sqrt{2\mu E}$, $\xi:=\sqrt{k^2-2\mu V_{0}}$, and the prime denotes differentiation with respect to the argument. \\
\quad\\
The bound-state wave functions are given by
\begin{equation}
\langle \vec{r}|E_{B}\ell m\rangle = i^{\ell+2}\kappa_{B}\mathcal{N}_{B}\left(\theta(d-r)h_{\ell}^{+}(i\kappa_{B}d)\frac{j_{\ell}(\xi_{B}r)}{j_{\ell}(\xi_{B}d)} + \theta(r-d)h_{\ell}^{+}(i\kappa_{B}r)\right)\mathcal{Y}_{\ell m}(\theta,\phi)\,,
\end{equation}
where $i\kappa_{B}$ is a bound-state pole of $f_{\ell}(k)$ on the positive imaginary $k$-axis, which satisfies
\begin{equation}
\xi_{B}\frac{j_{\ell}'(\xi_{B}d)}{j_{\ell}(\xi_{B}d)} = i\kappa_{B}\frac{(h^{+}_{\ell})'(i\kappa_{B}d)}{h^{+}_{\ell}(i\kappa_{B}d)}\,,\quad \xi_{B} = \sqrt{-\kappa_{B}^2-2\mu V_{0}}\,,\quad\mathrm{and}\quad
\end{equation}
\begin{equation}
\mathcal{N}_{B} = \frac{\xi_{B}}{\kappa_{B}d}\sqrt{\frac{2\kappa_{B}}{\kappa_{B}^2+\xi_{B}^2}}\biggl[ \kappa_{B}d \left|h_{\ell+1}^{+}(i\kappa_{B}d)\right|^2-(2\ell+1)\left|h_{\ell}^{+}(i\kappa_{B}d)h_{\ell+1}^{+}(i\kappa_{B}d)\right|\biggr]^{-\frac{1}{2}}\,.
\end{equation}

\end{appendix}


\end{document}